\documentclass[twocolumn,pre,aps,superbib,tightenlines,floatfix,superscriptaddress,showpacs]{revtex4}
\usepackage[T1]{fontenc}
\usepackage[latin9]{inputenc}
\usepackage{amsmath}
\usepackage{graphicx}
\usepackage{amssymb}
\usepackage{hyperref}

\begin{document}

\title{Solving mazes with memristors: a massively-parallel approach}
\author{Yuriy V. Pershin}
\email{pershin@physics.sc.edu}
\affiliation{Department of Physics and Astronomy and USC
Nanocenter, University of South Carolina, Columbia, SC 29208, USA}
\author{Massimiliano Di Ventra}
\email{diventra@physics.ucsd.edu}
\affiliation{Department of Physics, University of California San Diego, La Jolla, California 92093-0319, USA}

\begin{abstract}
Solving mazes is not just a fun pastime: they are prototype models in several areas of science and technology. However, when maze complexity increases their solution becomes cumbersome and very time consuming. Here, we show that a network of memristors - resistors with memory - can solve such a non-trivial problem quite easily. In particular, maze solving by the network of memristors occurs in a massively parallel fashion since all memristors in the network participate simultaneously in the calculation. The result of the calculation is then recorded into the memristors' states, and can be used and/or recovered at a later time. Furthermore, the network of memristors finds all possible solutions in multiple-solution mazes, and sorts out the solution paths according to their length. Our results demonstrate not only the first application of memristive networks to the field of massively-parallel computing, but also a novel algorithm to solve mazes which could find applications in different fields.
\end{abstract}

\maketitle

\section{Introduction} \label{sec1}

Mazes are a class of graphical puzzles in which, given an entrance point, one has to find the exit via an intricate succession of paths, with the majority leading to a dead end, and only one, or few, correctly "solving" the puzzle. Mazes - sometimes also called labyrinths - have been known since ancient times, the oldest presumably being the one created by Daedalus in Crete, as passed on by Greek mythology a few thousand years ago. They are used as prototype models in graph theory, topology, robotics, traffic optimization, psychology, and in many other areas of science and technology \cite{Pellow85a,Modesti98a,Teroa06a,Nakagaki00a,Crowe00a,Nelson04a,Blum97a,Reiners07a}. The ability to solve a maze is particularly important for transportation and robot control systems, e.g., to find the shortest path from a given point to another. In neuroscience, different maze solving algorithms could be compared with the approaches used by humans to solve mazes. In this way we could better understand the functioning of the human brain and advance new modeling of neural processes.

Algorithms to solve mazes vary from the simplest - and extremely slow - "random mouse" to mathematical search algorithms that operate on a sequential fashion to find the exit. However, all these methods suffer from very slow solution times when the complexity of the maze increases, with solution times sometimes increasing dramatically with increasing local connectivity of the network. Moreover, some researchers have demonstrated that certain biological and chemical systems can solve mazes \cite{Nakagaki00a,Lagzi10a}. For instance, in Ref. \cite{Nakagaki00a} an unexpected   behavior of a primitive organism was revealed by showing that an amoeba finds the minimum-length path between two points in a labyrinth connecting separate food sources. Such methods, however, are also slow and do not seem to be suitable for large mazes with complex connectivity.

In this paper, we suggest and demonstrate a new strategy for solving mazes that is instead based on massively-parallel computation as afforded by a network of memristors (short for memory resistors) \cite{chua71a}. Memristors are resistors whose resistance depends on the state history of the system, and can therefore record their past dynamics. These systems, which belong to the larger class of memory circuit elements - that includes also memcapacitors and meminductors \cite{diventra09a} - are attracting considerable attention due to their usefulness in diverse research areas \cite{pershin11a} ranging from memories \cite{Burr08a,Dietrich07a} to neuromorphic computing and learning \cite{pershin09b,pershin10c,jo10a}.

A network of memristors, complemented by some other standard electronic elements (such as field effect transistors), forms a {\it memristor processor} that, at the present level of technology, can be fabricated experimentally. In this work, instead, we use numerical simulations to explicitly show that a memristor processor is able to solve mazes. Most importantly, unlike existing approaches, the memristor processor requires only one step to find the maze solution. We emphasize though that the type of  massive parallelism used in the memristive processor can be thought of as {\it analog parallelism}. It is essentially different from that used in conventional computers, and rather shows some similarity with massively parallel computing with organic layers demonstrated recently \cite{Bandyopadhyay10a}.

It is also worth noting here that we could perform maze solving with a network of meminductors or memcapacitors. We choose to work with memristors since they are the most studied so far, and can be easily realized experimentally, thus allowing a practical and straightforward implementation of our algorithm.

\begin{figure*}[tb]
\begin{center}
\includegraphics[angle=0,width=12cm]{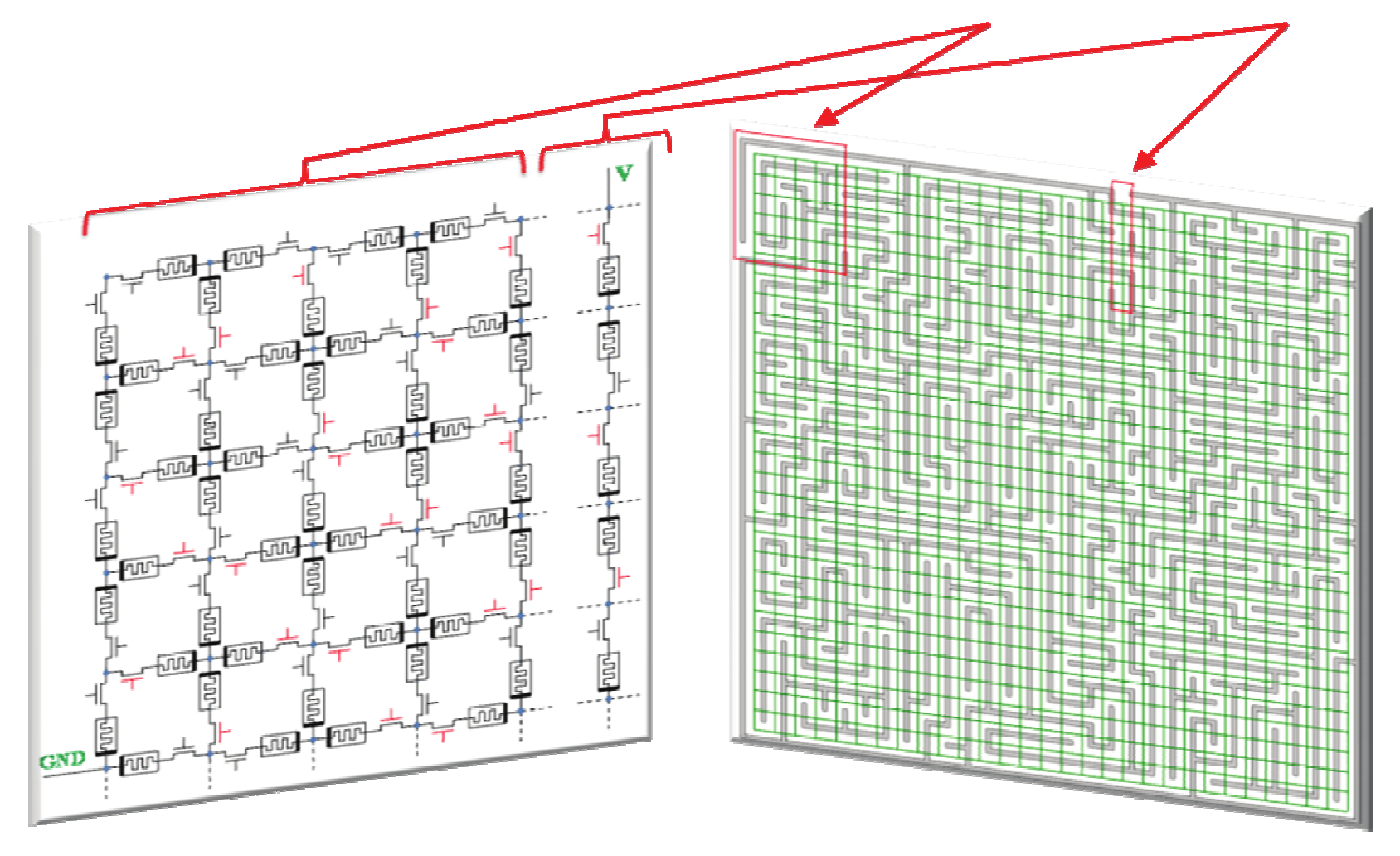}
\caption{(Color online) Maze mapping into a network of memristors (a.k.a. memristive processor). {\it Right panel}. The maze is covered by an array of vertical and horizontal lines having the periodicity of the maze. {\it Left panel}. Architecture of the network of memristors in which each crossing between vertical and horizontal lines in the array (in the right panel) is represented by a grid point to which several basic units consisting of memristors and switches (field-effect transistors) are linked in series. The maze topology is encoded into the state of the switches such that if the short line segment connecting neighboring crossing points in the array crosses the maze wall then the state of the corresponding switch is "not connected" (shown in red). All other switches are in the "connected" state. The external voltage ($V$) is applied across the connection points corresponding to the entrance ($V$) and exit (ground, $GND$) points of the maze. }
\label{fig1}
\end{center}
\end{figure*}

The remainder of this paper is organized as follows. In Sec. \ref{sec2}, we introduce the main part of our proposal, namely, the memristive processor. We discuss all important operation details of the memristive processor including its initialization, maze mapping and, finally, finding the maze solution. Sec. \ref{sec3} illustrates dynamics of maze solving by the memristive processor. In this section, we first introduce a mathematical model of the unit part of the processor, the memristor. Numerical simulations of solutions of both single-path and multiple-path mazes
by the memristive processor are presented. It is important to note that in the case of multiple-path mazes, the memristive processor not only finds all solutions but also sorts them out according to their length. We conclude with some final remarks and considerations on the proposed approach.

\section{Memristive processor} \label{sec2}

Let us then start by mapping a given maze into a network of memristors. This is shown in Figure \ref{fig1}. First of all, we superimpose  periodical arrays of vertical and horizontal lines on the maze. The period of this array corresponds to the intrinsic period of the maze (for non-periodic mazes, the period of line arrays should be selected in such a way to take into account all important maze features). The crossing points of vertical and horizontal lines define grid points of the memristive network. The network consists of basic units (memristors plus switches) connecting grid points. Since the direction of current flow in the network is not known a priori, the polarity of adjacent memristors (indicated by the black thick line in the memristor symbol in Figure \ref{fig1}) is chosen to be alternating. It is assumed that external signals can be applied to any grid points for the purpose of initializing the memristors' states as well as to read the calculation results. The externally-controlled switches are used to define the topology of the maze: a maze wall is modeled by a switch in the "not-connected" state. Such an architecture allows modeling different mazes on the same memristive processor without the need to fabricate a specific processor for each maze.

\begin{figure}[tbp]
\begin{center}
\includegraphics[angle=0,width=6cm]{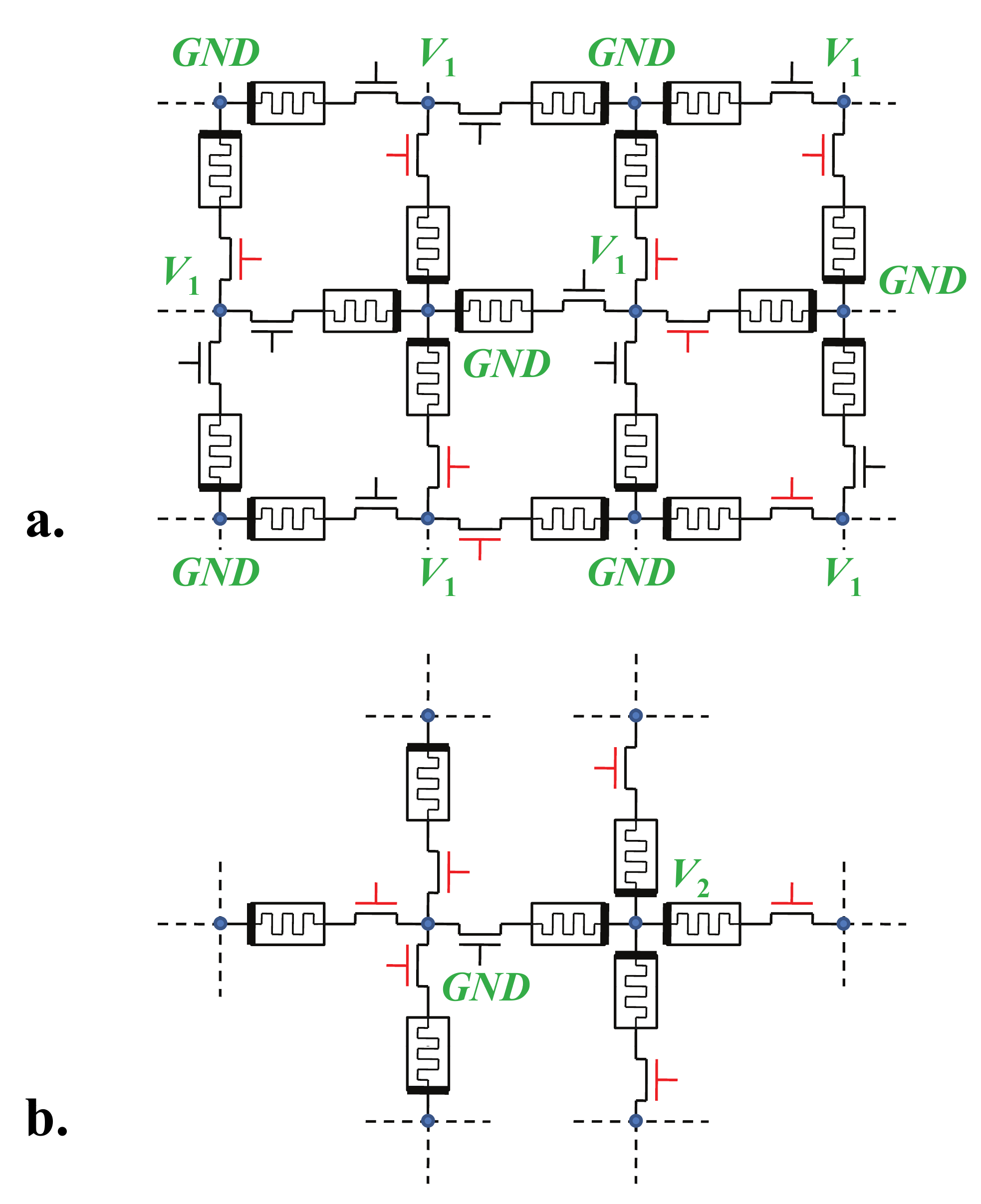}
\caption{(Color online) a) Initialization of the network of memristors can be performed by a simultaneous application of $GND$ and appropriately selected $V_1$ voltages in a chessboard-like pattern to all grid points for a sufficiently long period of time to securely switch memristors to the "OFF" state. Note that since the memristors' polarities are alternated in the network, the corresponding voltage polarities are also alternated. During the initialization, all switches should be in the "connected" state or should already encode the maze (as shown on the plot). b) In order to read a memristor's state, such a memristor (shown in the center) can be disconnected from the rest of the circuit putting the corresponding switches to the "not-connected" state (shown in read) and its own switch to the "connected" state and tested by the application of a short small-amplitude voltage pulse (or double bi-polar voltage pulse to minimize the disturbance of the memristor's state) of an amplitude $V_2$.}
\label{fig12}
\end{center}
\end{figure}

The processor initialization can be done by simultaneous application of $GND$, $V_1$ voltages in a chessboard-like pattern to all grid points of the memristive network (see Fig. \ref{fig12}a). The calculation performed by this memristive processor occurs when a constant voltage V is applied across the two grid points corresponding to the entrance and exit points of the maze as shown in Figure \ref{fig1}. In this case, the current flows only along those memristors that connect the entrance and exit points. The state of these memristors is changed by the current, thus the maze is solved in a massively parallel way, since all memristors in the network participate simultaneously in the calculation. Specifically, assuming that by the initial moment of time all memristors were initialized in the high resistance ("OFF") state, as time passes, every other memristor along the solution path changes its resistance, eventually switching into the low-resistance ("ON") state. Therefore, the chain of memristors in the "ON" state (or in an intermediate state if memristors did not have enough time to completely switch to the "ON" state) connecting the entrance and exit points represents the maze's solution. A possible approach to read the calculation result is described in Fig. \ref{fig12}b.

Furthermore, the states of the memristors connecting the entrance and exit points represent a solution at any given finite time after the initial one. If there are multiple paths, then the shortest one would contain less memristors and thus offers less resistance than the longest one, with all intermediate paths (in terms of length) offering a proportionate resistance. Therefore, since current flows in inverse proportion to the resistance of a path, at any given time, the change of state of a given path is proportional to the current in the path. The different paths of the maze can then be identified by the different state their memristors have during (or after) the switching process. In the next section, we explicitly discuss this solution feature. Here, we only point out that the memristive processor can be reduced to a network of standard resistors and, in principle, this could also solve a maze. However, in this case, additional external memory and local voltage measurement hardware will be required as the resistive network does not store the calculation results. In addition, in the case of a standard resistive network, previous calculation  results can not be easily used in subsequent calculations. Finally, a network of memristors is a complex non-linear dynamical system that can be potentially used to solve a larger class of problems, of which optimization ones
- as those described in this paper - are only a small class.

\begin{figure}[bpt]
\begin{center}
\includegraphics[angle=0,width=8cm]{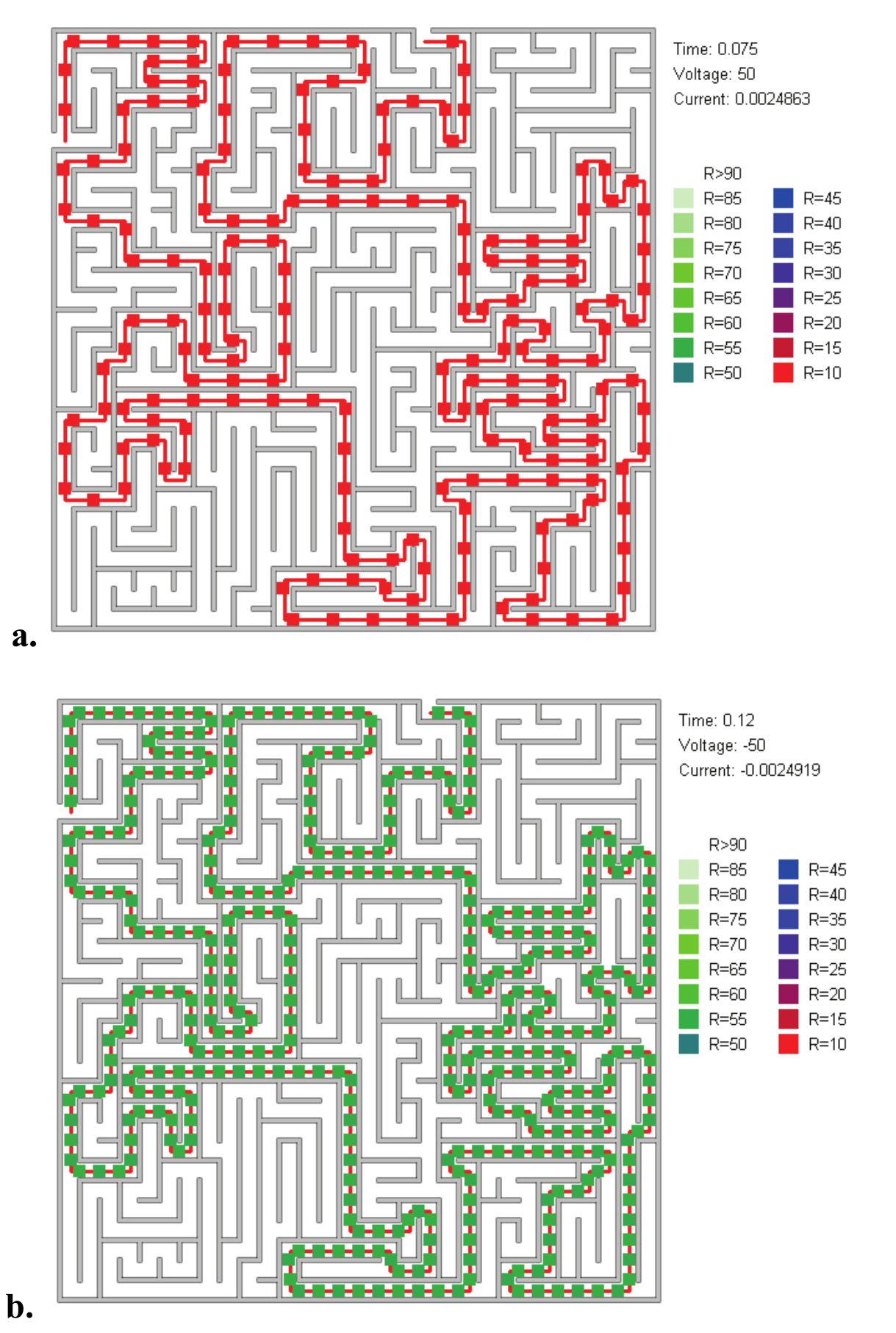}
\caption{(Color online) Solution of a single-path maze. a, Network state at $t=0.075$s. The chain of memristors in the low-memristance state (shown by red dots) clearly connects the entrance and exit points of the maze (note that memristors in the "OFF" state - when $R_{ij}^{M}(t=0)>90$ Ohms - are not shown). Here, every other memristor along the solution path is in the low-memristance state. b, Network state at $t=0.12$s. Note that at $t=0.1$s the sign of the applied voltage has been  changed. At  $t=0.12$s, each memristor along the solution path shows the maze solution. The resistance is in Ohms, the voltage is in Volts, and the current is in Amperes.}
\label{fig2}
\end{center}
\end{figure}

Experimentally, the suggested network could be fabricated using, e.g., CMOL (Cmos+MOLecular-scale devices) architecture \cite{Likharev05a} combining a single memristor layer with a conventional CMOS (complementary metal-oxide-semiconductor) layer. It was demonstrated in the past that many different classes of materials and systems exhibit memristive behavior including binary oxides
(TiO$_2$, CuO, NiO, CoO, Fe$_2$O$_3$, MoO, VO$_2$)~\cite{yang08a,inoue08a,lee07a,seo04a,driscoll09a,driscoll09b},
perovskite-type oxides (Pr$_{1-x}$Ca$_x$MnO$_3$, SrTiO$_3$:Cr)~\cite{asamitsu97a,fors05a,kim06a,meijer05a,nian07a},
sulfides (Cu$_2$S,Ag$_2$S)~\cite{terabe05a,tamura06a,waser07a}, semiconductors (Si, GaAs, ZnSe-Ge)~\cite{jo08a,dong08a,jo09a},
spintronics materials~\cite{pershin08a,pershin09a,wang09a} and organics~\cite{stewart04a,lai05a,alibart10a}. Moreover, hybrid memristor-CMOS  integrated circuits for reconfigurable logic applications as well as memristive memory chips combining memristive materials with transistors were recently developed experimentally \cite{xia09a,Dietrich07a}. Therefore, fabrication of a memristive processor is possible
at the current level of technology. A small-scale version of a memristive processor can be built using memristor emulators \cite{pershin10d}.

\section{Numerical simulations} \label{sec3}

\subsection{Model}

 In this section, we present numerical simulations of the dynamics of memristive processors. Numerical modeling of memristive networks is easily implemented and thus offers by itself a practical computational algorithm for maze solving. This computational approach, however, requires multiple computational steps (see below) as opposed to the real memristive processor discussed in the previous section which needs only a single step to perform the whole computation.

For the sake of clarity, we use a simple model of memristor \cite{Strukov08a} whose memristance (memory resistance) is given by
\begin{equation}
R_{ij}^M=R_{ON}x_{ij}+R_{OFF}\left( 1 -x_{ij}\right),\label{eq1}
\end{equation}
where $R_{ON}$ and $R_{OFF}$  are minimal and maximal values of memristance, $x_{ij}$ is the dimensionless internal state variable bound to the region  $0\leq x_{ij} \leq 1$, and ($i,j$) are grid indexes of a memristor to identify its location in the network. Similarly to Ref. \onlinecite{Strukov08a}, we choose the dynamics of $x_{ij}$  to be given by
\begin{equation}
\frac{\textnormal{d} x_{ij}}{\textnormal{d} t}=\alpha I_{ij}(t) \label{eq2},
\end{equation}
where $\alpha$ is a constant and $I_{ij}(t)$  is the current flowing through the memristor ($ij$).  At each time step, the potential at all grid points is found as a solution of Kirchhoff's current law equations obtained using a sparse matrix technique. The corresponding change in the memristors' states was computed using Eq. (\ref{eq2}).

\begin{figure}[t]
\begin{center}
\includegraphics[angle=0,width=7.3cm]{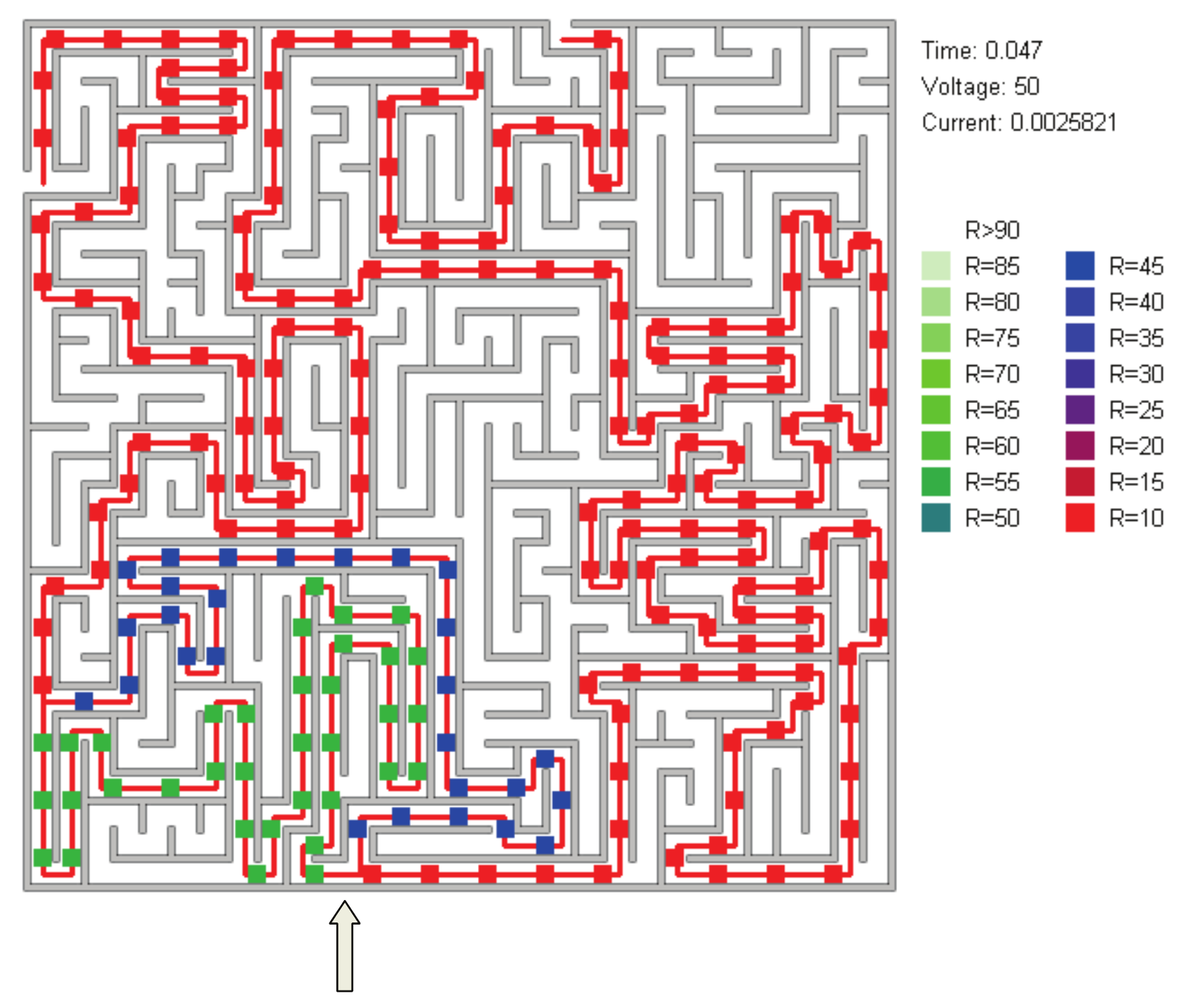}
\caption{(Color online) Solution of a multiple-path maze. Network state at $t=0.047$ s. The maze solution contains two common segments (red dots), and two alternative segments of different lengths close to the left bottom corner (blue and green dots). The memristance in the shorter segment (blue dots) is smaller than that in the longer segment (green dots) since the current through the shorter segment is larger and, consequently, the change of the memristors' state along this segment is larger. The arrow at the bottom indicates where a modification of the maze from Fig. \ref{fig2} has been made (position of the removed segment of the wall). The resistance is in Ohms, the voltage is in Volts and the current is in Amperes.}
\label{fig3}
\end{center}
\end{figure}

All numerical results reported in this paper were obtained using model parameters $R_{ON}=10$ Ohms, $R_{OFF}=100$ Ohms, $R_{ij}^M(t=0)=91$ Ohms, $x(t=0)=0.1$  and $\alpha=10^4$/(s$\cdot A$) for all memristors. The applied voltage was selected to be equal to 50V during the first 0.1 s time interval, and -50V at $t>0.1$s. The sign of the applied voltage was changed in order to better represent the maze solution as discussed below. Two types of mazes were considered: a single-path maze, and a multiple- (two-) path maze.

\subsection{Results}

We have applied the computational scheme described above to several mazes. In all cases, the correct maze solutions were found. Examples of our calculations are shown in Figures \ref{fig2} and \ref{fig3}. These mazes have been mapped on a $n \times n$ square memristive processor with $n=30$. Here, $n^2$ gives the total number of grid points. It is evident that the square geometry selected for this particular case is not generally required. Mazes of any shape (e.g., rectangular, circular, irregular) can be mapped on a square memristive processor of sufficient size.

Figure \ref{fig2} presents results of solution of a single-path maze (see also Supplementary Movie 1). We have found that with these parameters it takes approximately 0.06s to switch every other memristor along the solution path into the low-resistance "ON" state. The resulting sequence of memristors in the low-resistance state represents the maze solution as shown in Figure \ref{fig2}a. This maze solution can be better seen if we change the sign of the applied voltage and wait some time. Then, the resistance of memristors - along the solution path - that are in the "ON" state will increase, and the resistance of memristors in the "OFF" state will decrease. Figure \ref{fig2}b captures a specific moment of time when these memristances are equal. At this moment of time, every memristor along the solution path is in the same intermediate state and thus the solution of the maze is better visible.

A two-path maze whose solution is given in Figure \ref{fig3} was obtained from the maze shown in Figure \ref{fig2} by removing a single segment of the wall (its location is shown by an arrow in Fig. \ref{fig3}). In this maze, the current flowing through a common segment (red dots in Figure \ref{fig3}) splits between two possible paths according to their resistances. Therefore, memristors in the common segment change their resistance faster than those in the two possible paths (see Figure \ref{fig3} and also Supplementary Movie 2). Moreover, comparing memristors from two possible paths, memristors in the shorter path change their resistance faster than those in the longer path. As discussed previously, this allows sorting all possible solutions according to their length in such a way that the resistances of the memristors along shorter paths are smaller.

We note that the type of memristors and simulation parameters (such as $\alpha$ and $V$) have been selected for the sole purpose of demonstration of our idea. The resulting switching time (of the order of 0.05s, see the above discussion) is much longer than the typical switching times of nanoscale memristive devices that can be as short as 5ns (see, for example, parameters of nanoionic resistive random access memories in Refs. \cite{ITRS09a,pershin11a}). Since  with an appropriate choice of voltage/current magnitude the switching time of all the memristors along the solution path is on the order of the switching time of a single memristor, we can argue that the minimum time required to solve a maze of arbitrary complexity by our method can be as short as few nanoseconds or even less, since, as discussed above, full switching is not required to find a solution. Moreover, only one step (a single voltage pulse) is needed to find the solution. Therefore, the approach we suggest is {\it a priori} more efficient than any multi-step algorithm of present use. In addition, in simulations with memristors described by Eqs. (\ref{eq1})-(\ref{eq2}) we have always observed the system evolution toward a unique (for a given maze) stable solution. Therefore, the success rate of correct solution is 100$\%$ in our scheme.

Regarding the time complexity of the numerical modeling algorithm, it is mainly determined by the solution of a set of linear Kirchhoff's current law equations. The best theoretical estimate for a linear system  (the Coppersmith-Winograd algorithm \cite{Coppersmith90a}) is  O$(n^{2.376})$. This estimate thus provides the theoretical time complexity of the numerical maze solving algorithm suggested in this paper. We note that the time complexity of the breadth-first search algorithm that can be used to find the shortest path is only slightly better. In particular, it is given by O$(V+E)\sim \textnormal{O}(n^{2})$, where $V$ and $E$ is the number of vertices and edges, respectively \cite{Cormen09a}.
However, we would like to emphasize once more that the {\it hardware} implementation of the memristive processor requires a single computational step, and thus outperforms all existing maze solving algorithms.

\section{Conclusion} \label{sec4}

In conclusion we have shown how a network of memristors - a memristive processor - can solve mazes in a massively-parallel way. This approach can be realized experimentally with available systems and devices, or simply implemented on a computer. The hardware implementation of the memristive processor is superior to any existing maze solving methods and therefore it is ideal when the complexity of the maze increases with increasing local connectivity of the graph. Although we have considered a processor based on memristors, a network of memcapacitors or meminductors \cite{diventra09a} can also be used for  massively parallel calculations. As of now, several experimental systems exhibiting memcapacitive and meminductive properties are known \cite{pershin11a}. Electronically, memcapacitive and meminductive circuits can be emulated using memristors \cite{pershin09e,pershin11b}.

Moreover, we anticipate that the memristive processor can facilitate the solution of - or solve - many other computational problems. Examples of such problems include the traveling salesman problem, graph theory problems, etc. The memristive processor can then be used as a complete computational device, or as a supplemental tool for traditional computing hardware.
Since memristors (as well as memcapacitors and meminductors) are generally asymmetric devices, unidirectional graphs can also be easily realized on appropriate memristive processors. We thus envision their use in a large set of applications in both basic science and technology.

\section*{Acknowledgement}

M.D. acknowledges partial support from the NSF Grant No. DMR-0802830.

\bibliography{maze}

\newpage

\noindent {\bf Supplementary materials} to the manuscript {\it Solving mazes with memristors: a massively-parallel approach} by Y. V. Pershin and M. Di Ventra

\vspace{0.5cm}

\noindent Supplementary Movie 1

\vspace{0.2cm}

\noindent This movie shows dynamics of maze solving for a single-path maze. The time is in seconds, resistance is in Ohms, voltage is in Volts, and current is in Amperes. The red line links network grid points that are not separated by maze walls and are connected by memristors whose $R_{ij}^M<80$ Ohms.  Note that at $t=0.1$s the polarity of applied voltage changes.

\vspace{0.7cm}

\noindent Supplementary Movie 2

\noindent This movie shows dynamics of maze solving for a multiple-path maze. The multiple-path maze (see Fig. 3 of the article) is obtained from the single-path maze (Fig. 2 of the article) by removal of a single wall segment (see Fig. 3 for details). The time is in seconds, resistance is in Ohms, voltage is in Volts, and current is in Amperes. The red line links network grid points that are not separated by maze walls and are connected by memristors whose $R_{ij}^M<80$ Ohms. Note that at $t=0.1$s the polarity of applied voltage changes.

\vspace{0.5cm}

\noindent The movies can be found at \href{http://www.physics.sc.edu/~pershin/memdevices.htm}{http://www.physics.sc.edu/~pershin/memdevices.htm}.

\end{document}